\documentclass[screen,sigconf]{acmart}

\newcommand{\system}{RevTogether}

\AtBeginDocument{%
  }

\copyrightyear{2025}
\setcopyright{cc}
\acmConference[]{}{}{}
\acmDOI{}
\acmISBN{}

\begin{document}

\title{RevTogether: Supporting Science Story Revision with Multiple AI Agents}

\author{Yu Zhang}
\email{yui.zhang@my.cityu.edu.hk}
\orcid{0000-0002-8574-111X}
\affiliation{%
  \department{Department of Computer Science}
  \institution{City University of Hong Kong}
  \city{Hong Kong}
  \country{Hong Kong}
}

\author{Kexue Fu}
\email{kexuefu2-c@my.cityu.edu.hk}
\affiliation{
  \department{School of Creative Media}
  \institution{City University of Hong Kong}
  \city{Hong Kong}
  \country{Hong Kong}
}

\author{Zhicong Lu}
\email{zlu6@gmu.edu}
\affiliation{%
  \department{Department of Computer Science}
  \institution{George Mason University}
  \city{Fairfax}
  \state{Virginia}
  \country{USA}
}

\begin{abstract}
As a popular form of science communication, science stories attract readers because they combine engaging narratives with comprehensible scientific knowledge. However, crafting such stories requires substantial skill and effort, as writers must navigate complex scientific concepts and transform them into coherent and accessible narratives tailored to audiences with varying levels of scientific literacy. 
To address the challenge, we propose \system{}, a multi-agent system (MAS) designed to support revision of science stories with human-like AI agents (using GPT-4o). \system{} allows AI agents to simulate affects in addition to providing comments and writing suggestions, while offering varying degrees of user agency.
Our preliminary user study with non-expert writers (N=3) highlighted the need for transparency in AI agents’ decision-making processes to support learning and suggested that emotional interactions could enhance human-AI collaboration in science storytelling.
\end{abstract}



\begin{teaserfigure}
  \includegraphics[width=\textwidth]{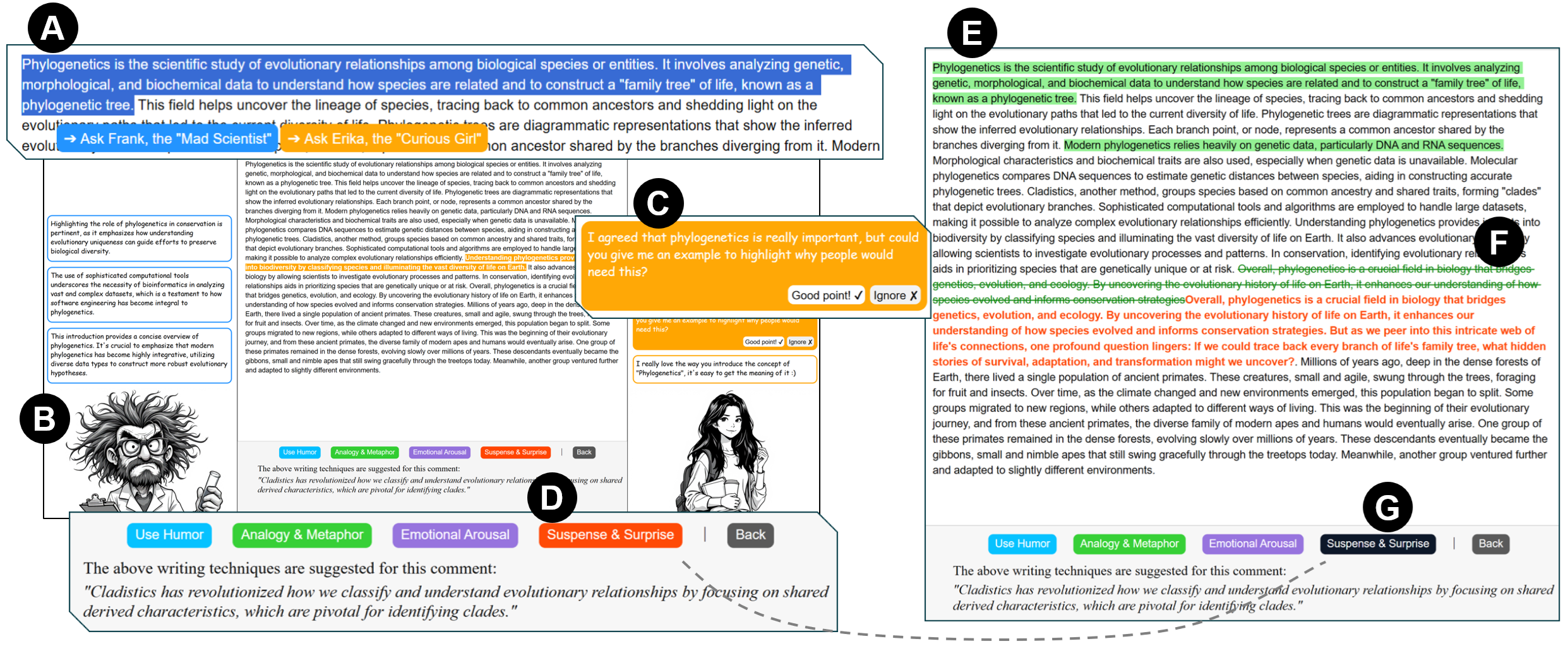}
    \caption{Overview of the \system{} System.
     (A) The writer can ask the commentator agent for feedback on a randomly selected segment of the texts. (B) The commentator agents present human-like avatars with emotional appearances matching the sentiment of comments. (C) The writer can review and respond to each comment. (D) Once the writer agrees with a comment, a writing assistant agent will provide writing technique suggestions for addressing the problems mentioned in comments. (G) The writer can further choose a technique to see (E) which parts of the texts are suggested to apply the technique and (F) review detailed revision suggestions provided by AI agents.}
  \Description{Overview of the \system{} System.
     (A) The writer can ask the commentator agent for feedback on a randomly selected segment of the texts. (B) The commentator agents present human-like avatars with emotional appearances matching the sentiment of comments. (C) The writer can review and respond to each comment. (D) Once the writer agrees with a comment, a writing assistant agent will provide writing technique suggestions for addressing the problems mentioned in comments. (G) The writer can further choose a technique to see (E) which parts of the texts are suggested to apply the technique and (F) review detailed revision suggestions provided by AI agents.}
  \label{fig:teaser}
\end{teaserfigure}

\maketitle

\section{Introduction}
Science stories have long captivated broad audiences with varying levels of scientific literacy by blending engagement, entertainment, and comprehensible scientific knowledge ~\cite{davies2019science, joubert2019storytelling, zhang2023understanding}.
However, crafting quality stories from complex scientific facts highly relies on writers' expertise and mastery of advanced writing strategies ~\cite{elshafie2018making, august2020writing, he2024engage}. 
For example, even scientists, who are usually the primary source of new scientific discoveries, also struggle to write engaging and accessible narratives to share their research insights with the general public ~\cite{tuttle2023promoting, cote2018scientists, smith2018breaking}. Fortunately, advanced computing and artificial intelligence (AI) technologies could bring opportunities to lower the barriers to science story creation, empowering a wider range of individuals to participate in science communication.

With the rapid development of Large Language Models (LLMs) ~\cite{brown2020language} and Generative Artificial Intelligence (GenAI) ~\cite{muller2022genaichi}, researchers in multiple disciplines including Human-computer Interaction (HCI) have noticed the potentials of applying AI to support various tasks in science communication ~\cite{alvarez2024science, biyela2024generative, nishal2024understanding, williams2022hci}, including writing for audiences from the general public ~\cite{jiang2024llm, gero2023social, schafer2024trust, tatalovic2018ai}. Previous work has explored approaches to scaffolding the creation of science news and social media content using AI ~\cite{kim2023metaphorian, gero2022sparks, laban2024beyond}. For example, LLMs are used to apply particular writing strategies (e.g., ``hook'' in Long et al.~\cite{long2023tweetorial} and ``metaphor'' in Kim et al.~\cite{kim2023metaphorian}), provide creative inspirations ~\cite{gero2022sparks}, and retrieve additional information from external sources  ~\cite{laban2024beyond}. 
However, these approaches neglect feedback from other perspectives, focusing solely on fulfilling the writer’s demands.
Feedback from different perspectives is essential in writing because it helps to build an iterative revision process for writers to continuously improve the outcomes ~\cite{benharrak2024writer, du2022understanding, nelson2009nature}. Lacking feedback prevents writers from making changes with clear directional guidance ~\cite{long2023tweetorial, wigglesworth2012role}, especially for science stories that can have a wide range of readers from scientists to the general public ~\cite{elshafie2018making, burns2003science}.

Inspired by Benharrak et al.’s previous research that highlights the potential of social factors (e.g., relation, valence) in designing AI personas for writing feedback ~\cite{benharrak2024writer}, 
we propose \system{}, a multi-agent system (MAS) that supports writers in revising science stories together with two human-like commentator AI agents and a writing assistant agent.
Specifically, we designed the personas of the two commentator agents as a ``mad scientist'' and a ``curious girl'', representing the two sides of the ``dialogue model'' in science communication---scientists and ordinary people ~\cite{trench2008towards, reincke2020deficit}---for providing feedback in diverse perspectives. 
To enhance emotional expression, we introduced character avatars with dynamic emotional responses for the commentator agents in \system{}, providing non-verbal feedback to engage the writers.


Through a preliminary user study with three non-expert writers, we highlighted the importance of the transparency of AI agent's suggestion generation process because it could be helpful for writers who want to learn from AI. This finding aligned with the previous work on human-AI alignment in writing tasks ~\cite{wang2025jumpstarter, gero2023social, goyal2024designing, chakrabarty2024can}, revealing \emph{learning} as a new demand in human-AI collaborative writing. 
Besides, we also found that emotional reactions simulated by AI agents could serve as effective feedback to stimulate writers' thoughts in science storytelling.
Our findings could inspire future research on AI agent design for broad tasks in content creation in the context of science communication.

\section{System Design and Implementation}
\system{} aims to support science story revision with on-demand feedback provided by two commentator agents and a writing assistant agent.
To begin with, we chose the Human-AI teaming (HAT) analysis framework ~\cite{newendorp2024apple, sepich2021human} as design guidelines to operationalize our idea into a feasible prototype system.
According to the HAT framework, \system{} is a \emph{multi-agent-single-human} system in which the two commentator agents and the writing assistant agent collaborate and interact with a single human writer. With different roles played and tasks performed by the agents, \system{} offers multiple levels of user agency.

The overall layout of \system{}'s user interface (UI) is similar to common document editing tools (e.g., Microsoft Word and Google Docs) where the centric area is used for text editing and the side columns are used for supportive information such as commentary and the character avatars of commentator agents (\autoref{fig:teaser}).
We design the UI more like an editing tool rather than chatbot is because we hope to highlight science story writing as the main task. Direct conversation with agents are intentionally not supported as it may interrupt writers' thought processes.


\begin{figure*}[hbt]
    \centering
    \includegraphics[width=15cm]{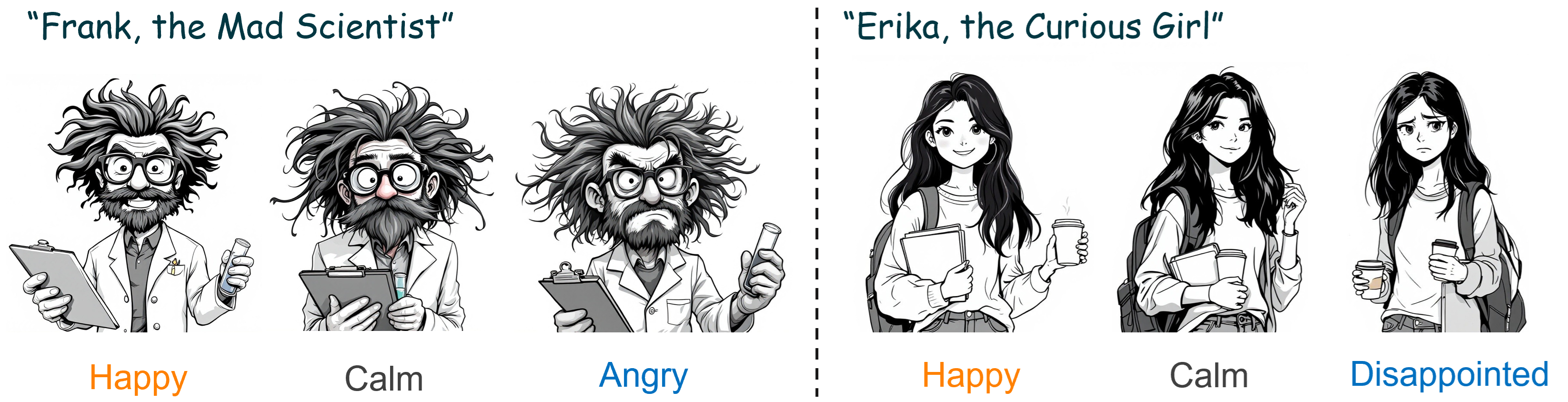}
    \caption{Character Avatars with different affects for Commentator Agents. Each character has three avatars with positive, neutral, and negative affect respectively. They share ``happy'' and ``calm'' for positive and neutral affects, while the ``mad scientist'' character has ``angry'' while the ``curious girl'' character has ``disappointed'' as negative affect.}
    \Description{Character Avatars with different affects for Commentator Agents. Each character has three avatars with positive, neutral, and negative affect respectively. They share ``happy'' and ``calm'' for positive and neutral affects, while the ``mad scientist'' character has ``angry'' while the ``curious girl'' character has ``disappointed'' as negative affect.}
    \label{fig:avatars}
\end{figure*}

\subsection{Designing the Commentator Agents}
In order to simulate feedback from potential readers with different perspectives, we design and implement two commentator agents that role-play a ``mad scientist'' and a ``curious girl''. The two specific personas are derived from the two types of major stakeholders (i.e., scientists and the general public), as well as typical readers of science stories, in science communication.
Based on the HAT framework, the commentator agents are \emph{critic} type ~\cite{sepich2021human} of agents who take the responsibilities for giving feedback on demand.

\subsubsection{Asking for Comments}
When editing a science story using \system{}, the writer can ask commentator agents to comment on user-specified segments (e.g., sentences, paragraphs, and full texts) of the manuscript. Precisely, when selecting a piece of texts, a floating menu will appear beneath the selection area to let the writer decide whom to ask for (\autoref{fig:teaser} (A)).
Once the writer chooses either of the commentators, the corresponding agent will take the selected texts as a focus to make a comment. 
Newly-generated comments will appear above the agent's character avatars next to the editing area (\autoref{fig:teaser} (B)). Multiple comments are placed in a bottom-up direction from the latest to the earliest.
When hovering mouse on a piece of comment, the writer can either accept (clicking on ``Good point!'') or reject (clicking on ``ignore'') it (\autoref{fig:teaser} (C)). 
The actions will trigger agents to respond with emotional reactions.


\subsubsection{Emotional Reactions}
\label{sec_emotional_reactions}
To establish the emotional engagement between the writer and commentator agents by leveraging the human-likeness perspective in the HAT framework ~\cite{bonny2024increasing, kim2023anthropomorphic}, we introduced comic-style character avatars for the two commentators displaying beside the editing area in \system{} (\autoref{fig:teaser} (B)). 

Particularly, for each commentator we created three avatars in different affects (emotional states): positive (i.e., happy), neutral (i.e., calm), and negative (i.e., angry or disappointed). Thus, we use the avatars as a non-verbal cue to convey emotional reactions (\autoref{fig:avatars}).
When asked to generate comments, we require the LLM to indicate the sentiment of each comment. The sentiment value can be positive, neutral, or negative. 
With the sentiment information associated with each comment, while the writer hovers the mouse over a comment for reviewing, the commentator character will change to the avatar of the corresponding affect.
While the writer not hovering or there is not comment yet, the character will remain in the neutral avatar by default.

Besides reflecting on the sentiment of comments, we additionally allow the commentator characters to respond to the writer's acceptance/rejection actions with affect changing actions. 
If the writer accepts a comment, the character will turn to the positive avatar staying for one second and turn back. The action performs like the avatar smiles to the writer as a compliment. On the contrary, rejection action will triggers the negative avatar as a mplain. The actions are designed to be in a non-verbal way to avoid interruptions.

\subsection{Designing the Writing Assistant Agent}
While human-like comments can provide some clues for how the manuscript could be changed, the writer still needs to figure out specific writing strategies for improvement. 
To provide clearer guidance of composing compelling science stories, we introduce the writing assistant agent. Unlike the commentators, the assistant agent is designed to be of \emph{companion} type ~\cite{sepich2021human}, that is less active and passively responds to the writer's requests to provide suggestions with particular writing techniques. Therefore, the writing assistant agent does not have an avatar as it does not need to provide human-like comments and conveys emotional reactions to the writer.

The assistant agent will only be activated when the writer explicitly accepts a comment made by the commentator agent. Once a comment is accepted, the suggestions will be displayed below the editing box as a series of tags (\autoref{fig:teaser} (D)) indicating which writing techniques can be used for addressing the comment. 
When the writer clicks on a tag, the agent will highlight all the places that could be changed by adding a green background to the texts, giving visual hints to the writer for further editing (\autoref{fig:teaser} (E)). 
Although each comment is associated with a piece of user-specified text, the writing suggestions will still consider the full story. Because only make changes in the specified texts could be limited and break the consistency through the entire story. 

In addition to highlighting the places for changing, the assistant agent also provides an AI-revised version of each highlighted place. Once the writer clicks on a highlighted segment, the auto-revised version will be displayed inline using a red color and a weighted font (\autoref{fig:teaser} (F)). The writer can easily adopt the auto-revised version by double clicking on it.

\begin{table*}[hbt]
  \caption{Techniques for Science Story Writing. We use four techniques in \system{}: (1) Humor, (2) Analogy and Metaphor, (3) Emotional Arousal, and (4) Suspense and Surprise.}
  \Description{Techniques for Science Story Writing. We use four techniques in \system{}: (1) Humor, (2) Analogy and Metaphor, (3) Emotional Arousal, and (4) Suspense and Surprise.}
  \label{tab:strategies}
  \begin{tabular}{p{2cm}p{5.75cm}p{5.75cm}p{1.5cm}}
    \toprule
        Technique & Definition & Purpose & Source \\
    \midrule
        Humor & The use of wit, jokes, or light-hearted language to make complex topics more engaging and enjoyable. & (1) Capture attention, (2) Simplify understanding, (3) Make the content relatable to readers. & ~\cite{yeo2020scientists, yeo2021emotion} \\
        
        Analogy and Metaphor & Compare complex ideas to familiar concepts to enhance understanding. & (1) Simplify obscure topics by relating them to everyday experiences or imagery, (2) Make the content memorable. & ~\cite{august2020writing, kim2023metaphorian} \\

        Emotional Arousal & The use of evocative language or storytelling to trigger readers' emotions and create a deeper connection & (1) Engage readers, (2) Make the content memorable, (3) Inspire curiosity. & ~\cite{davies2019science, yeo2021emotion, zhang2023understanding} \\

        Suspense and Surprise & Build anticipation through uncertainty and captivate readers by delivering unexpected twists or revelations. & (1) Engage readers, (2) Stimulating curiosity, (3) Make the content memorable. & ~\cite{august2020writing, huang2020good, zhang2023understanding} \\

    \bottomrule
\end{tabular}
\end{table*}

\subsubsection{Embedding Writing Techniques in Agent's Suggestions}
To incorporate effective feedback for revision, we analyzed the common techniques in science story writing from the literature and embedded these techniques into prompts so that the writing assistant agent can make suggestions with those concrete techniques.
In this way the agent can reduce the chance to give verbose or unspecific feedback ~\cite{benharrak2024writer}.
Since the writing techniques and strategies of science stories refers to an endless space of creativity, we only adopted four well-established techniques commonly mentioned in previous research ~\cite{yeo2020scientists, yeo2021emotion, august2020writing, kim2023metaphorian, davies2019science, zhang2023understanding, huang2020good} in \system{} at this moment. 
Specifically, we summarize and present the detailed definitions, purposes, and literature sources for those techniques (\autoref{tab:strategies}).


\subsection{Supporting Revision in Different Levels of User Agency}
Using \system{}, the writer can always take the initiative to freely edit the manuscript, while \system{} enables an additive AI-supported iterative revision process, allowing the writer to receive supportive information on three levels of user agency (\autoref{fig:workflow}).

\subsubsection{Human-like comments for high agency}
(\autoref{fig:workflow} (A))
The human-like comments made by commentator agents leave high agency to users. The commentator agents will not interfere when editing texts, but will only share subjective feedback like humans. The writer can use the provided comments at will to figure out a strategy to improve the draft.

\subsubsection{Suggestions on writing techniques for medium agency}
(\autoref{fig:workflow} (B))
While the clues obtained from subjective comments are usually ambiguous and unclear, 
the writing assistant agent thus is introduced to provide suggestions about specific writing techniques.
First, when the writer explicitly accepts a comment, the writing assistant agent will reveal which techniques are possibly useful for addressing the issues mentioned by the comment (\autoref{fig:teaser} (D)). At this stage, the agent still does not edit any texts but provide clearer directions for the writer to think about and move forward. This leaves a medium level of user agency in the human-AI collaboration in \system{}.

\subsubsection{Suggestions on revisions for low agency}
\begin{figure*}[htbp]
    \centering
    \includegraphics[width=15cm]{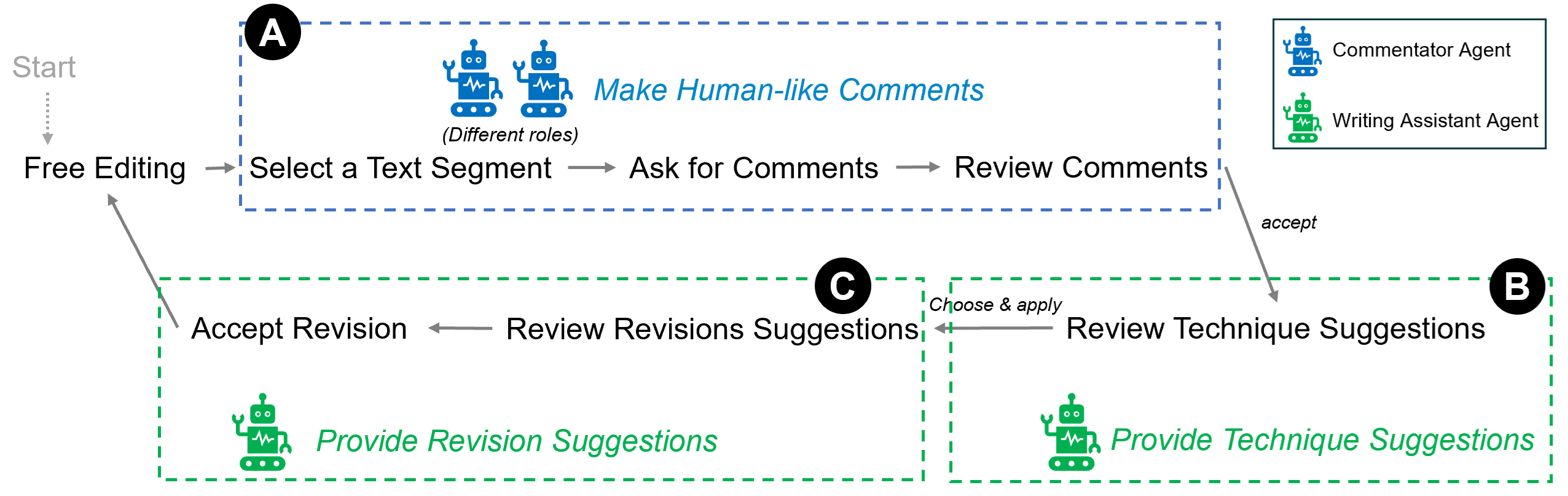}
    \caption{The Iterative Revision Workflow. \system{} enables an AI-supported iterative revision process, which starts from free editing and allows the writer to receive supportive information on three levels of user agency throughout the process: (A) Human-like comments with high agency; (B) Writing technique suggestions with medium agency; and (C) Detailed Revision suggestions with low agency.}
    \Description{The Iterative Revision Workflow. \system{} enables an AI-supported iterative revision process, which starts from free editing and allows the writer to receive supportive information on three levels of user agency throughout the process: (A) Human-like comments with high agency; (B) Writing technique suggestions with medium agency; and (C) Detailed Revision suggestions with low agency.}
    \label{fig:workflow}
\end{figure*}

(\autoref{fig:workflow} (C))
Once the user chooses to let the agent to provide revision suggestions (\autoref{fig:teaser} (G)), the agent will then give hints on which parts of the texts the writing technique can apply to (\autoref{fig:teaser} (E)). Only when the writer clicks on a hint will the system completely show an AI-revised version of texts (\autoref{fig:teaser} (F)) allowing the writer to simply accept by clicking on the revised texts (shown in red color).
The revision suggestion feature of writing assistant agent leaves the lowest user agency in \system{} as AI takes the control of editing and directly involved in story generation.

\subsection{Implementation}
We built \system{} as a Web application using Flask \footnote{https://flask.palletsprojects.com/en/stable/} on Python for its back-end and React \footnote{https://react.dev/} for the front-end. Particularly, we employ react-quill \footnote{https://github.com/zenoamaro/react-quill} to implement the text editor for achieving customized text selection, highlighting and editing. 
For the large language model (LLM) used to power the AI agents, we adopted GPT-4o \footnote{https://openai.com/index/hello-gpt-4o/} offered by a commercial cloud service provider that is accessible and stable in our region.
All the features of the AI agents in \system{} are achieved by our customized prompts and the conversations between the back-end and the remote LLM, where the back-end is responsible for assembling dynamic parameters (e.g., user-selected texts) from the front-end into predefined prompt templates and process the result from the remote LLM. 
Particularly, the avatars of the commentator agents are manually generated using the FLUX-pro-1.1 \footnote{https://www.fluxpro.ai/flux-1.1-pro} model.

\section{Preliminary User Study}
To explore the potential of human-AI collaborative writing in science story creation with \system{}, we conducted a preliminary user study with 3 participants (1 female, 2 male, aged between 22 to 30) who are all PhD students from our local university to evaluate \system{}.
The study results highlight the importance of the transparency of AI agent's suggestion generation and the effects of emotional reactions simulated by AI agents in \system{}.
These findings provide design implications for future iterations of \system{} and reveal opportunities for future research on human-AI collaborative content creation in science communication. 

\subsection{Procedure}
For each participant, the study lasted for about 60 minutes and the session was conducted in a one-on-one manner with the first author of this paper. 
The procedure started from signing the consent (\textasciitilde 5 min), getting familiar with \system{}, and knowing the backgrounds and motivations of building \system{} (\textasciitilde 10 min).
Next, we required the participant to generate a short story based on an academic paper they like, which can also be one of their own published papers, using GPT-4o with a fixed prompt given by us (only replacing the PDF file uploaded to the conversation).
Using this story as an initial draft, each participant was instructed to request comments from both commentator agents at least twice and to accept at least one comment from each for further interaction with the writing assistant.
The ultimate goal is to compose a better science story with \system{}. During the editing process, we encouraged the participants to think-aloud about their experience using \system{}. The editing process lasted about 30 minutes, and after which we conducted a short post-study interview with each participant, asking them to reflect on their thoughts and the most impressive experience with \system{}. 
The entire study procedure was screen-recorded and audio-recorded for analysis.

\subsection{Findings and Future Directions}
Given the exploratory nature of our study and the limited sample size of three participants (referred to as P1, P2, and P3), our findings do not allow for definitive conclusions regarding the effectiveness of our design. Instead, we focus on reporting the key insights that emerged from the preliminary user study. These findings highlight potential areas of interest and provide a foundation for future research directions to further investigate and refine \system{}. 

\subsubsection{Highlighting the Transparency of the Decision-making Process of AI Agents}
Explicitly showing a ``thought process'' of AI agents can help writers make sense of AI-generated suggestions and learn from the suggestions to improve their own skills in science story writing.
When asked about system features to extend in post-study interview, P1 asked ``\textit{Can the system show me how AI reads the story line by line to decide which parts of the texts are applicable for adding humor? Because I want to know the thoughts behind a suggestion to determine if it is reasonable or just an illusion of the language model behind.}''. P1 believed a visible step-by-step process showing how AI agent comes up with a writing suggestion would help him to learn from AI in science story writing, and also make him more confident with his later decision to either accept or reject an revision offered by the writing assistant agent. 
P3 also mentioned about the idea of learning from AI through a more transparent process of obtaining suggestions from the writing assistant agent. ``\textit{By using the system for writing such a story, I always want to have improvement on my own writing skills rather than letting AI write for me, otherwise why didn't I just let GPT generate full stories and keep improving by itself? I hope to see more details and reasons behind to have deeper understandings of the writing techniques}'', said P3.

In summary, we found that more information regarding how and why AI makes a suggestion (i.e., the transparency of decision-making of AI agents) need to be exposed in the interactions between the human writer and AI agents, so that the writer can learn from and rely on this information to make their own decisions effectively.
This finding echoes previous research on human-AI alignment in AI-supported writing systems ~\cite{wang2025jumpstarter, goyal2024designing, chakrabarty2024can, gero2023social}, showing that allowing users to better learn from AI should also be a design consideration for building AI agents for writing.


\subsubsection{The Effects of Emotional Expressions of AI agents in Science Story Writing}
Although our participants knew that the emotional reactions of AI agents' avatars were simulated by programmed rules and LLMs rather than real human emotional expressions, they could still feel strong feedback from the changing face expressions of the avatars, which is a kind of non-verbal cue in human-AI communication ~\cite{newn2020nonverbal}.
P3 mentioned, \textit{``Seeing the agent appears unhappy, I felt like I was motivated to carefully re-consider the comment again. This reminds me that my story could be read by people with very different opinions and might affect their emotions.''}
All of the three participants agreed that the emotional reactions as feedback affected their decision making during story editing. \textit{``It's really interesting and funny to see the avatar change faces when accepting or rejecting a comment''}, P2 commented.

Previous research have revealed the importance of emotion in science communication in terms of how emotion of the audiences could be affected and considered in design and content creation to enhance the public engagement of science ~\cite{lin2012role, trench2021rethinking, huang2020good, zhang2023understanding}.
However, how the emotion of content creators (e.g. science story writers) and the simulated ``emotion'' of AI agents can affect science communication content creation, remains under-explored. Future research in this line may extend the knowledge of the effects of emotion in human-AI co-creation in the contexts of science communication.

\section{Limitations and Future Work}
While our preliminary user study demonstrates the potential of \system{} in supporting science story revision, there are several limitations to address. First, the small sample size of three participants, all of whom were non-expert writers, limits the generalizability of our findings. 
Secondly, while emotional expressions of AI agents were perceived by the participants in the preliminary study, their actual impact on user engagement and human-AI collaboration was not systematically examined, leaving their effectiveness as an open question.

In future work, we plan to expand the scale of our user study by involving a larger and more diverse group of participants to obtain a more comprehensive and robust evaluation of the system's usefulness. Additionally, we aim to invite professional science story writers to assess \system{}, providing expert perspectives on its design and functionality to further refine the system. Furthermore, we intend to conduct ablation studies to investigate whether the current emotional expressions of AI agents effectively elicit user responses and enhance their collaboration and engagement with the agents. These efforts will help us gain deeper insights into \system{}'s impact and identify areas for improvement to better support human-AI collaboration in science storytelling.

\section{Conclusion}
In this paper, we propose \system{}, a multi-agent system (MAS) that facilitates the iterative revision process in science story creation by employing Large Language Model (LLM) -based AI agents to provide feedback and suggestions with different levels of user agency. Through a preliminary user study with three non-expert writers, we revealed the potential of \system{} and future research directions regarding understanding and supporting human-AI collaboration in writing tasks within the contexts of scientific communication.

\bibliographystyle{ACM-Reference-Format}
\bibliography{main}


\end{document}